# Modelling colossal magnetoresistance manganites

**T. V. Ramakrishnan**

Department of Physics, Banaras Hindu University, Varanasi-221005, India

Email: tvrama@bhu.ac.in

**Abstract**. Rare earth manganites (with alkaline earth ions partially substituting them), i.e. $Re_{1-x}Ak_xMnO_3$, are being intensively explored for the last decade or more because of the promise of magnetoelectronic applications as well as because of complex and unusual phenomena in which electronic, structural and magnetic effects are intertwined. A brief survey of these and a description of the three strong local interactions of the $e_g$ electrons (in two different orbital states at each site), namely with Jahn-Teller phonon modes (strength $g$), with resident $t_{2g}$ spins (ferromagnetic Hund's rule coupling $J_H$) and amongst each other (the Mott Hubbard correlation U) form the background against which efforts at modelling manganite behaviour are described. A new two fluid model of nearly localized $l$ polarons and band ($b$) electrons for low energy behaviour emerges for large $g$; some of its applications are mentioned here. First I describe some results of large U, $J_H$ calculations in single site DMFT (Dynamical Mean Field Theory) which includes the effect of all the strong local correlations. These results are directly appropriate for the orbital liquid regime, found typically for $0.2<x<0.5$, and not too low temperatures. We show that many characteristic manganite phenomena such as an insulating ferromagnetic ground state, thermal insulator metal transition (nearly coincident with the paramagnetic to ferromagnetic transition), colossal magnetoresistance (cmr), materials systematics dependent on the specific Re and Ak ions and the observed low effective carrier density can all be understood qualitatively as well as quantitatively. We also discuss the two 'phase' coexistence frequently found in these systems, and show that electrostatic coulomb interactions mute $lb$ phase separation into nanoscale electronic inhomogeneity with $l$ regions and $b$ puddles. Finally, some problems of current interest as well as general ones arising, eg polarons and the physics of large electron phonon coupling $g$ in the adiabatic regime, are mentioned.

## 1. Introduction

The occurrence of colossal magnetoresistance or cmr in alkaline earth (Ak) doped rare earth (Re) manganites ($Re_{1-x}Ak_xMnO_3$ with $0<x<1$) more than a decade ago [1] has led to widespread exploration of these compounds which were discovered by Jonker and van Santen [2] in the nineteen fifties and studied by them. While the initial impetus was the possibility of magnetic devices based on cmr and related activity continues, it quickly became clear that this family is home to very diverse, unusual and poorly understood phenomena as well as phases. These solid state oxides have emerged as second only to the high $T_c$ cuprates in the overall global level of research activity; both present fundamental challenges to our understanding of how electrons behave in solids. A number of articles [3–6] and books [7–10] review this field.

*1.1. Phenomena*
The diversity of phenomena in doped manganites {eg [3-10] and references there} is suggested

(figure 1) by the phase diagram of $La_{1-x}Ca_xMnO_3$ [11], a well studied member of the family. Various kinds of antiferromagnetic insulating phases (e.g. A , Neel, and CE), metallic as well as insulating ferromagnetic states, orbital long order with or without antiferromagnetism, charge order, insulator metal transitions over a broad range of doping (0.2<x<0.5), long range cooperative antiferro-order of Jahn-Teller distorted octahedra for small x, are all seen to be present. The insulator metal transition is nearly coincident with the paramagnetic to ferromagnetic or Curie transition (temperature $T_c$), implying that electronic and magnetic effects are strongly coupled. The connection between structural and electronic properties is illustrated by the fact that while the charge and orbitally ordered structure (with x~0.5) forms a superlattice and is electrically an insulator, a relatively small magnetic field destroys this superstructure ('melts' it), and the resulting system is metallic. Other manganites show characteristic similarities and differences. For example in $La_{1-x}Sr_xMnO_3$, the Curie transition is from a metal to a metal, whereas in $Pr_{1-x}Ca_xMnO_3$ it is best described as insulator to insulator. The transitions between phases are generally of second order. There are however several cases of first order transition, e.g. the magnetic field induced melting of charge order. Rather than describe further the large number of phenomena in these compounds [Refs. 3-10 give a good idea of these], I mention below three of their striking mutually related general physical characteristics. I do not discuss in detail any specific region (eg the fascinating half doped regime where charge and orbital order are endemic) or a class of effects, but try to focus on common phenomena, basic interactions and models of general significance.

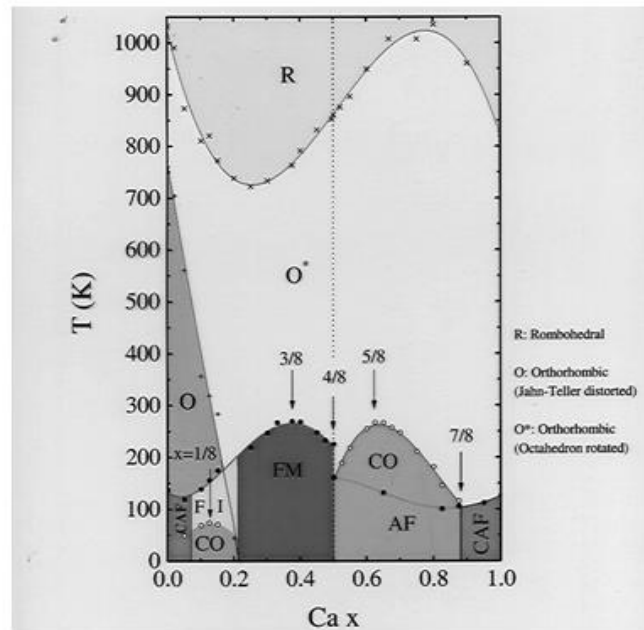

Figure 1: The phase diagram of $La_{1-x}Mn_xO_3$ in the doping x and temperature T plane. The structural phases shown are rhombohedral R, orthorhombic O* with rotated octahedra but without long range Jahn-Teller distortion order, and orthorhombic O with rotated octahedra and long range J-T order. CAF (canted antiferromagnetic), FI (ferromagnetic insulator), FM (ferromagnetic metal),CO (charge /orbitally ordered),and AF (Neel antiferromagnetic) regions are also shown.{ Adapted from [ 11 ] }.

One general characteristic is the persistent proximity of metallic and insulating states. Normally, a system is either a metal or an insulator depending on whether electronic states near the Fermi energy are extended or localized (as they are in Anderson, Mott or band insulators). Consequently, a transition from one to the other occurs only under special conditions of pressure, temperature or composition. However, as can be seen for example from figure 1, LCMO has an insulator to metal transition over the **entire** doping range 0.2<x<0.4 as the temperature is lowered below about 200 -300 K. This obviously implies that current carrying extended states and 'insulating' localized states are close to each other in energy over a wide range of doping conditions in this compound, as they are for many others in this family.

Another general characteristic is the extreme sensitivity of physical properties of manganites to small perturbations. Colossal magnetoresistance (see e.g. figure 2) is perhaps the best known example. Near the Curie transition, the resistivity of most manganites decreases enormously (the fractional change being of order unity) in an external magnetic field of order a few Tesla; the change is two to three orders of magnitude larger than that in a typical metal, where it is generally due to the effect of the magnetic field on the electron trajectory, and is therefore determined by the factor $(\omega_c\tau)^2$, $\omega_c$ being the cyclotron frequency and $\tau$ the electron relaxation time. The large isotope effect is another instance; the Curie temperature depends strongly on the oxygen isotopic mass, being for example about 10% higher in LCMO with $O^{16}$ than with $O^{18}$ [12]. Perhaps the ultimate in isotope effects is the observation [13] that the ground state of $(La_{1-y} Pr_y)_{0.7} Ca_{0.3}MnO_3$ for y = 0.75 is a metal with $O^{16}$ and an insulator with $O^{18}$. Finally, the insulating charge/orbitally ordered state (near x~0.5) with $T_{CO}$~250 K 'melts' to a metal on application of a magnetic field typically of order 10 Tesla; the Zeeman energy associated with the field is clearly more than an order of magnitude smaller than the charge ordering energy scale.

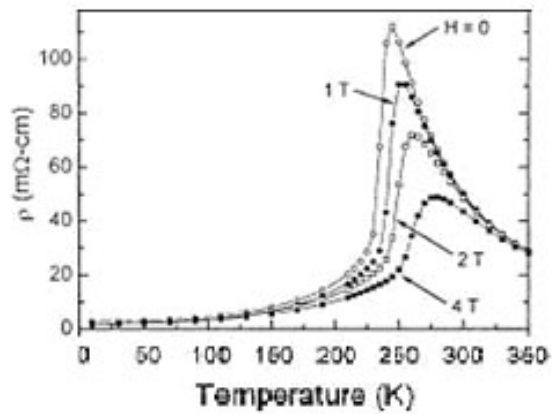

Figure 2: Colossal magnetoresistance as observed in a $La_{0.75}Ca_{0.25}MnO_3$ sample with a Curie temperature $T_c$ of about 230K. The resistivity is shown as a function of temperature for various values of the field. The change in resistance with magnetic field is seen to be colossal near $T_c$.

Thirdly, there is overwhelming evidence [5,9] for the simultaneous presence in manganites of two regions or kinds of states on spatial scales varying from nanometres [14] to microns [15] and timescales ranging from less than $10^{-13}$ seconds [16-18] to about $10^{-6}$ seconds [19] to $10^0$ seconds (static). One of these regions is lattice distorted and insulating while the other has no lattice distortions and is metallic. Often, these regions are referred to as phases, even though they can be nanometres in size and not macroscopic (i.e. not in the thermodynamic limit). Whether such a two 'phase' coexistence is intrinsic to manganites or indeed to all strongly correlated electronic systems (eg stripes in cuprates [20] and metal/insulator droplets in 2DEG [21]), being their defining characteristic and related to their 'electronic softness' [22], or whether it is extrinsic is a question of great current interest in the physics of strongly correlated electron systems [22]. I briefly discuss it later in Sections III and IV.

*1.2 Interactions at Work*

Structurally, the manganites are slightly distorted perovskites $ABO_3$ which can be viewed as corner sharing octahedra with the Mn (or B) ions at the centre of the $BO_6$ octahedron, the O ions at the corners, and the A ions in the interstitial space between octahedra. There is a well known ideal size for the A ions, and deviation from it in the mean as well as local fluctuations in it due to A site doping have many interesting and systematic consequences.

I describe now the electronic degrees of freedom relevant for low energy properties of the manganites, as a preliminary to modelling the low energy ones. In the $MnO_6$ octahedron at each lattice site, the s p orbitals of Mn and O are strongly bonded, so that the associated charge transfer levels are well below the Fermi energy. In the octahedral environment of the manganites, the *d* levels of the Mn ions split into lowlying threefold degenerate $t_{2g}$ states and higher energy twofold degenerate $e_g$ states separated by about 2 eV, the crystal field energy. In $LaMnO_3$ the four 3d electrons of $Mn^{3+}$ are in a high spin (S = 2) state; the interorbital coulomb repulsion is large and one is in the weak crystal field regime. The three $t_{2g}$ electrons have a total spin S= (3/2) and are (effectively) ferromagnetically coupled with the $e_g$ electron spin via $J_H$, the Hund's rule coupling, leading to a total spin S=2. $J_H$ is estimated to be large, about 2 to 3 eV. On doping with divalent alkaline earths which substitute for the trivalent rare earths, a fraction x of sites has $Mn^{4+}$ ions. At every site then, whether occupied by $Mn^{3+}$ (four 3d electrons, probability (1-x)) or by $Mn^{4+}$ (three 3d electrons, probability x), there are always three $t_{2g}$ electrons, so that their only relevant low energy degree of freedom is spin (S=(3/2)), labelled by the spin operator **$S_i$** at site i. The low energy electronic behaviour is thus determined by the $e_g$ electrons; the ingredients governing their dynamics are mentioned now. An average fraction (1-x) of the Mn lattice sites is occupied by them in the doped compound, with a fraction x of sites having no $e_g$ electrons.

The two degenerate $e_g$ orbitals ($d_{x2-y2}$ and $d_{3z2-r2}$) constitute the explicit electronic degrees of freedom at each site. These are often formally described as two components ($\alpha,\beta$) of a pseudospin (S =1/2), the corresponding Pauli pseudospin operator being labeled by $\tau$ (with components $\tau_x, \tau_y, \tau_z$). In a tight binding d electron model, the nearest neighbour hopping amplitude for the $e_g$ electron is parametrized by a single number t; the dependence on the orbital states (initial and final) as well as on the spatial direction is determined by the d orbital wavefunction symmetry, and is given by the matrices $\mathbf{A}_v$ {Slater Koster factors, eg [23]} below, where v are the Cartesian directions x,y and z.

$$H_K = - (t/4) \sum_{<ij>} \mathbf{d}_{i\sigma}^+ \mathbf{A}_v \mathbf{d}_{j\sigma} \qquad (1)$$

In equation 1, $\mathbf{d}_{j\sigma}$ removes a d electron with spin $\sigma$ from the site states ($\alpha,\beta$) at site j. From the calculated electronic structure [24-30], a bandwidth 2W of order 2 to 4eV [24-27,29,30] is estimated. If parametrized by a tight binding band, as is done eg in [27,29,30] the hopping matrix element t is about 0.2 to 0.3 eV.

Such a model is a simplification and the hopping t represents a second order term, the primary step being charge transfer from (or to) an oxygen p orbital which is the intermediate state. An explicit dp or two band model has been discussed in detail {see eg [31] for the $LaMnO_3$ end}. Here I work with a one (d) band model as is common in the field. Some reasons supporting this simplification are the following. Firstly, the 'd' states are local symmetry adapted mixtures of d and p orbitals. Secondly, the integrated out 'p' state (or charge transfer band) energy $\varepsilon_p$ is well away from the Fermi level, so that for energies less than $|\varepsilon_p - \varepsilon_F|$, a picture which concentrates on locally d like states (around each Mn ion) is accurate. Thirdly, the strong interaction effects of Jahn Teller coupling as well as $J_H$ and U all involve the d state on site.

As mentioned above, one strong interaction present at each site i is effectively described as ferromagnetic Hund's rule coupling between $t_{2g}$ spins $S_i$ and $e_g$ spins $s_i$, namely

$$H_{Hund} = \sum_i H^i_{Hund} = -J_H \sum_i S_i \cdot s_i \qquad (2)$$

It was pointed out very early, by Zener [32] and more pertinently in the way used today, by Anderson and Hasegawa [33], that for large $J_H$ the effective nearest neighbour hopping amplitude for $e_g$ electrons depends on the angle between the corresponding $t_{2g}$ spins since the $e_g$ spin direction is enslaved to that of the $t_{2g}$ spin on site, and the $e_g$ electron conserves its spin direction on hopping. The hopping amplitude and thus the kinetic energy gain due to $e_g$ electron motion is maximum when the $t_{2g}$ spins point parallel to each other, i.e. are ferromagnetically aligned. This novel 'double exchange' ferromagnetic coupling is fundamental to manganites. It implies for example that metallicity (due to $e_g$ electron motion) and ferromagnetism are strongly connected with each other (see below, Section II, for details and a critique).

It is well known that in the cubic perovskite $ABO_3$ a quadrupolar ($l = 2$) Jahn Teller distortion of the $BO_6$ octahedron removes the twofold degeneracy of the $e_g$ levels and one of the resulting states is lower in energy. This is prominent in $LaMnO_3$ where below 780K the distortions order cooperatively ('antiferro' order) while above this temperature there is evidence from EXAFS measurements [34] that it is locally present. The interaction involved can be described as follows. The on site Jahn Teller coupling between the $e_g$ orbitals and lattice modes $\mathbf{Q_i} = Q_{xi}, Q_{zi}$ which are specific combinations of Mn-O bond length changes can be written following Kanamori [35] as $H^i_{JT} = g\, \mathbf{d}_{i\sigma}^+ \boldsymbol{\tau} \mathbf{d}_{i s} \cdot \mathbf{Q_i}$ with coupling strength $g$. The Hamiltonian is, explicitly,

$$H_{JT} = \sum_i H^i_{JT} = g \sum_i (d_{i\alpha_s}^+ d_{i\alpha\sigma} - d_{i\beta\sigma}^+ d_{i\beta\sigma}) Q_{zi} + g \sum_i (d_{i\alpha\sigma}^+ d_{i\beta\sigma} + d_{i\beta\sigma}^+ d_{i\alpha\sigma}) Q_{xi} \qquad (3a)$$

The lattice potential energy for a displacement $\mathbf{Q_i}$ is given in the harmonic approximation by $(1/2) K\, \mathbf{Q_i}^2$ where we have assumed the force constants K for the the two modes $Q_{ix}$ and $Q_{iz}$ to be the same, which is nearly correct [36]. The lattice or phonon Hamiltonian (for the relevant Einstein like modes) is

$$H_{ph} = \sum_i H^i_{ph} = \sum_i \{V(\mathbf{Q_i}) + (1/2M) p_i^2\} \qquad (3b)$$

where $V(\mathbf{Q_i}) = (1/2) K \mathbf{Q_i}^2$ in the harmonic approximation, and the nearly equal reduced mass M of the modes is approximately that due to oxygen. The two energy eigenvalues of the Jahn Teller term are $\pm g |\mathbf{Q_i}|$ for an electron present at the site i. For the lower energy state, the total potential energy minimum of $(-g^2/2K) = -E_{JT}$ occurs at $|\mathbf{Q_i}| = Q_o = (g/K)$ (in the harmonic approximation). $E_{JT}$ is the gain in energy because of Jahn Teller distortion $Q_o$. For the other eigenvalue, the potential energy minimum is still at $|\mathbf{Q_i}| = 0$.[1]

The Jahn Teller mode frequency $\omega_{JT}$ ($= \sqrt{K/M}$) is believed to correspond to an energy of about 0.06 eV, based on the identification of certain Raman spectra peaks with the Jahn Teller modes [37]. The parameter which describes the strength of the electron lattice coupling is $g$. It is known to be large in manganites; the related Jahn Teller energy $E_{JT}$ lies in the range 0.5 to 2.0 eV [27,28,38]. This energy scale is inferred in a number of ways. One is to use the observed Jahn Teller distortion in $LaMnO_3$,

---

[1] We note that there is a continuous degeneracy of the eigenvalues; they are independent of the direction (given by the angle $\theta_i = \tan^{-1}(Q_{xi}/Q_{zi})$) in which the pseudomagnetic field at site i points, though the eigenstates (the orbital admixtures) depend on it. This is reduced to a discrete threefold degeneracy when cubic anharmonic terms in $Q_i$ are included [35]

and the measured JT mode frequency $\omega_{JT}$ along with the mode reduced mass M, and find $E_{JT}$ (or the electron lattice coupling $g$ and the force constant K) using the expressions above [38]. Another is to use results from band structure calculations, eg [27] or from cluster calculations. All of these result in an $E_{JT}$ of about 1 eV, the range being from 0.5 to 2 eV.

The electron phonon coupling $g$ is not expected to change much with doping x, for several reasons. Experimentally, short time scale experiments [16-18] show that the consequent Jahn Teller distortion does not change much with x even in LSMO, in the range 0.25 > x > 0.0. Its reduction by mobile hole screening is expected to be small both because the holes are not very mobile and because the coupling is local; the screening length is a few lattice constants (see eg figure 7). The electron phonon interaction is generally characterized by two dimensionless coupling constants, namely the strength $\lambda$ (= $g^2/Kt$) and the adiabaticity $\gamma$ (= $\hbar \omega_{ph}/t$). In manganites, $\lambda \sim 1$ to 4, and $\gamma \sim 0.2$ to 0.3.

The Jahn Teller distortion $Q_o$ in LaMnO$_3$ is large, about 0.15 A [39]. This has been known for a long time, and indeed some of the early theories of the 780K ordering in LaMnO$_3$ modelled it as an order disorder transition of the local Jahn-Teller distortion already present in each MnO$_6$ octahedron, and supported this hypothesis through the measured integrated entropy change connected with it [40]. The work of Kanamori [35] mentioned above describes the Jahn-Teller and $e_g$ electron degrees of freedom explicitly, and is the basis for studies of the undoped manganite. With the renewed recognition that large Hund's rule coupling effects such as double exchange are prominent in doped manganites, it began to be felt that the Jahn Teller effect is not essential for phenomena in them. In influential work, Millis, Littlewood and Shraiman [41] and Millis, Muller and Shraiman [42] pointed out that features such as the large resistivity of manganites near the Curie transition ($T_c$) and the large resistivity change there cannot be understood in terms of only double exchange (or spin) interactions, and that it is essential to include the effect of strong coupling to the lattice leading to polarons.

The third strong on site interaction is the Mott Hubbard repulsion. In general, there are several comparable terms here, corresponding to orbital and spin indices of the $e_g$ orbitals, eg [43]. Of these, given that for large $J_H$ the $e_g$ electron spin direction is enslaved to that of the $t_{2g}$ spin (so that there is no $e_g$ electron spin degree of freedom, effectively) the term of relevance is the repulsion U when the two $e_g$ electrons are in different orbitals. Its actual value is a subject of debate. General trends in perovskite transition metal oxides, photoemission experiments and cluster calculations [24,43,44] lead to values ranging from 5 to 10 eV. Distinctly smaller values, of order 2 - 3eV or so, have been argued for by Millis and others [45]. The value in relation to the bare bandwidth 2W (= 2zt in a nearest neighbour tight binding model) of the $e_g$ electrons is obviously an important basic question. The latter is estimated to be about 2 to 4 eV. Thus for the larger values of U ($\geq$ 5 eV) one is definitely in the strong correlation regime U >> zt, while for the smaller values, mean field approaches to the effect of U and consequent independent electron like theories may be adequate. Experimentally, LaMnO$_3$ above the 780K transition (in the phase with no magnetic or structural long range order) is an insulator, i.e. it is a Mott insulator. In tune with this, with the correlation energy systematics in such compounds (as inferred eg from core photoemission spectra), and the broad consensus in the field, I take U to be large, of order 5eV or more. The U term can be written

$$H_U = U \sum_i n_i (n_i - 1) \qquad (4)$$

Thus the total Hamiltonian of a collection of Mn $e_g$ electrons on a lattice (two orbitals per site) consists of intersite hopping given by Eq. 1, Hund's rule coupling Eq.2, Jahn Teller coupling as well as related phonons (Eq. 3) and finally Mott Hubbard correlation U (Eq.4). It is thus given by

$$H = H_K + H_{Hund} + H_{JT} + H_{ph} + H_U \qquad (5)$$

There are several additional interactions and physical parameters which may be significant in different contexts. One is strain, which seriously influences the energetics of correlated octahedral tilts and distortions, with the corner sharing constraint. A manifestation of this correlation is the 'antiferro' relation between the nearest neighbour JT distortions so that the local and global strain is small.. Such interactions have been discussed extensively [46]. In addition, there can be direct correlation between orbital states and between JT displacements on nearest neighbour sites [27]. There is also the octahedron breathing mode coupled to the total number of $e_g$ electrons on a site irrespective of its orbital distribution. One qualitative consequence of it might be the observed steep decrease in the JT ordering temperature with doping; the added holes induce breathing mode distortions which disorder the antiferro distortive arrangement of the Jahn Teller polarons [36]. The effect of long range coulomb interaction between the $e_g$ electrons and the Re, Ak ions is also ignored (see Section III and references there for a discussion of this).

Disorder has major effects on the observed properties of manganites [47]; that due to doping is nearly unavoidable. This is exemplified in the problem of A site disorder, inevitable unless the dopants are ordered as has been achieved in $La_{0.5}Ba_{0.5}MnO_3$ [47]. It is known that in case of this system, the physical properties, and indeed the phases, are very different depending on A site order [6,47,48]. One necessarily has ion size and charge variation on an atomic scale which has direct electronic effects. The ion size can also act via strain for example. Experimentally, two classes of well known consequences are the following: local ion size fluctuation, for a general incommensurate doping x, strongly reduces the Curie temperature $T_c$ [49]. Attfield and collaborators [49] found that $T_c$ decreases linearly (down to less than half its 'uniform' value) with the rms variation in ion radius, assuming that ions are distributed completely randomly on each site. Tokura and others have shown [6,47,48] that for a given composition x near (1/2), the relative sizes of the Re and Ak ions (both of which occupy the A site of the $ABO_3$ or perovskite structure) determine whether the low temperature phase is orbitally ordered or only magnetically and the nature of the phase above the ordering temperature. There are in addition effects due eg to deliberately introduced disorder, as with Al substituting randomly for Mn in $La_{1-x}Sr_xMnO_3$ [50], relatively long range (microns ?) strain randomness due to surfaces, interfaces, cracks, strain fields associated with small defects, etc.. The effects are large, and are clearly of a piece with the observed high sensitivity of physical properties to perturbations.

**2 Models and their consequences**

Given the complexity of the problem, theoretical models for manganites focus on one or more of the above interactions as crucial, and develop approximate theories whose results resemble some experimental features of manganites. We mention briefly here a few classes of theories, and then describe the microscopic two fluid (*lb*) model developed by us. The consequences of the latter are described in the next Section. It is broadly accepted that in addition to double exchange, electron phonon coupling is very significant. Indeed, manganites can be viewed as a 'laboratory for electron phonon physics'[39].

2.1 *Ab initio electronic structure*

The electronic, magnetic and structural properties of manganites have been actively explored for more than a decade using a variety of theoretical ab initio methods for electron dynamics in solids, eg Hartree Fock [51], LDA [25,27], LSD [24,26] without and with [28] self interaction,
LDA+U [27,29,30]. An ab initio approach for this family poses well known special challenges such as the following. These systems have (unfilled d shell) electrons with strong coulomb interactions rather than quasi independent ones for which conventional band theory is most reliable. There is a variety of phases with energies very close to each other. Nonzero temperature behaviour of doped nonstoichimetric disordered compounds is of interest often while most accurate calculations can be done for ordered stoichiometric compounds at T=0; the difference can be quite radical. In spite of these limitations, the calculations are essential for several reasons. They constitute the base of our theoretical understanding in many cases and it is through them that one arrives at values for the sizes of the necessary ingredients for a realistic model (eg interactions and Hamiltonian parameters). I illustrate these general comments now in the context of manganites.

All electronic structure calculations agree broadly with the picture described above of crystal field split $e_g$ and $t_{2g}$ states. The band of electronic states near the Fermi level is $e_g$ like, while the one identifiable with $t_{2g}$ states is well below the Fermi level as is the oxygen p band. There is clear evidence for large on site inter d- orbital coulomb repulsion leading to a large Hund's rule like ferromagnetic exchange coupling $J_H$. Further, cluster calculations, constrained electron occupation number estimations of U[25] and interpretation of photoemission experiments [44], all point to large U, in the range 5 to 10 eV while the bandwidth is 2 to 4 eV. The naturalness and the importance of Jahn Teller like distortions are also explicitly brought out.

The properties of the parent $LaMnO_3$ compound as obtained from various ab initio approaches are quite illuminating. Approximations which neglect U lead to either a metal, or to an AF insulator with a relatively small gap and a small J-T distortion $Q_o$. There is a synergistic effect between U and $Q_o$; on including the former, one obtains a phase similar in properties to that observed, with large distortion $Q_o$. With pressure, it is seen experimentally [52] that the JT distortion of the unit cell decreases continuously and vanishes at about 18GPa. However, the resistivity continues to be insulating, and becomes metallic only above 32GPa; this intervening phase could be a Mott insulator, not a band one. Detailed recent calculations (LDA+U, LDA+DMFT) [30] on the other hand suggest that this phase is likely to be an insulator because of as yet unobserved small long range distortions present in addition to large U. Such a conclusion points to a limitation of this kind of approach in directly confronting experiment. For example, the experiments could imply that at intermediate pressures there is a Jahn-Teller liquid phase with no long range order, but short range order, of the sort found at zero pressure but above the J-T ordering temperature [34]. Such a liquid phase is not realistically accessible to ab initio calculations.

The electronic structure results provide estimates for parameters for simplified models of the sort mainly discussed here. The numbers have been quoted earlier. Additionally, authors have developed and used tight binding as well as Jahn Teller coupling parametrizations to fit the ab initio results. Two examples are the use of a model very similar to Eq (6) in [29] and [30], and the exploration of cooperative Jahn Teller effects in [27].

2.2 *Strong Hund's rule coupling $J_H$ only, or Double Exchange*

As pointed out above, there is a strong effectively ferromagnetic coupling $J_H$ between $t_{2g}$ spins and $e_g$ electrons onsite. Thus when an $e_g$ electron hops from one site to the next and the associated $t_{2g}$ spins do not point in the same direction, the effective hopping amplitude is the bare one times the spin overlap factor because of the large $J_H$; the latter is cos $(\theta_{ij}/2)$ where $\theta_{ij}$ is the angle between the classical $t_{2g}$ spins at i and j [33]. This double exchange (DE) mechanism leads simultaneously to ferromagnetism and metallicity in doped manganites. Further, at a given doping, as temperature increases and $t_{2g}$ spins disorder thermally, the average kinetic energy of the $e_g$ electrons decreases and their incoherence increases. One thus expects a strong correlation between metallicity and the ferromagnetic-paramagnetic or Curie transition, which is observed. By the same token, application of a magnetic field near $T_c$ polarizes the $t_{2g}$ spins and should increase the $e_g$ electron mobility or current. For these qualitative reasons, it is generally believed that double exchange explains colossal magnetoresistance (in a generic doping regime) as well as the association of the insulator metal transition with the paramagnetic ferromagnetic transition. The most detailed early calculations in the double exchange model (which cannot be solved exactly) are those of Furukawa [53], who used single site dynamical mean field theory (DMFT), namely obtained selfconsistently the local magnetization and the time dependent onsite self energy of the $e_g$ electron, which can be done exactly for infinite dimensions [54].This nonperturbative method is known to be accurate for strong interaction phenomena in which spatial correlations do not play a critical role; it is expected to work well here in the absence of orbital order etc.. One finds a Curie transition, but to an incoherent metal, not an insulator. The magneoresistance can be large, but not of the colossal size often seen. There is a large body of work based on DE as the sole essential ingredient for manganite physics. This may be appropriate for some systems eg LSMO which are metallic above and below $T_c$ and for some doping regimes.

However, the pure double exchange model has several other well known shortcomings, a few of which are the following. Some doped manganites are insulating paramagnets with a Curie transition to a metal or to an insulator. Almost all manganites have, for low doping, an **insulating** ferromagnetic ground state. Now, in a DE model, the only generic way a doped manganite can avoid being metallic is via Anderson localization of electronic states near the Fermi energy, this arising from $e_g$ electron motion in a disordered background of $t_{2g}$ spins. Careful calculations [55] show that the fraction of states localized thus is very small; for example, at x ~ 0.25 or so, only about 0.5% of the $e_g$ states are localized with the maximum possible disorder of $t_{2g}$ spins. Secondly, Millis and coworkers [41,42] pointed out that the large changes in electrical resistivity observed near $T_c$ cannot be understood in a pure spin disorder picture and that the effect of strong electron lattice coupling leading to Jahn Teller polaron formation must be included. This is also supported by a large body of experimental evidence for polarons. Short and long range orbital order may require additional ingredients. There are also other spin interactions present, eg the superexchange between $t_{2g}$ spins which can be antiferromagnetic or even ferromagnetic, depending on the relevant $e_g$ orbital configurations, and a ferromagnetic virtual double exchange which is possible when a JT polaronic site and a hole are nearest neighbours { [56,57,59] and Section III below }. Competition between antiferromagnetism and DE ferromagnetic exchange, leading to possible phase separation when frustrated and amplified by disorder, has been suggested by Dagotto and coworkers (see eg [5,9]) to be the cause of the ubiquitous two 'phase' coexistence phenomenon in manganites. We [56-59] have argued that at least for x<0.4 the ferromagnetism observed in manganites arises overwhelmingly from virtual double exchange which is of second order in the hopping, ([56,57,59] as well as Eq.8 and the discussion there) rather than double exchange which is linear in it.

*2.3 Strong Hund's rule coupling $J_H$ and Jahn Teller interaction g*

As mentioned above, Millis and coworkers [41,42] showed that it is essential to include the effect of the strong coupling of the twofold degenerate $e_g$ orbitals to the octahedral symmetry breaking Jahn-Teller lattice modes. The static distortion induced by this coupling produces a polaron, lower in energy by $E_{JT}$. Using single site DMFT, they determined selfconsistently the on site polaronic distortion, the self energy of the $e_g$ electron and the mean field magnetization. In addition to such equilibrium quantities, transport properties were calculated. They found, for the 'half filled' case (ie n=1 or x=0) that at high temperatures, ie temperatures above $T_c$, there are large polaronic distortions which diminish and disappear below $T_c$. The (second order) Curie transition can be a metal insulator transition. This is suggested to be a general feature of manganites even away from x=0, eg doped ones. One feature of the calculation noted by the authors is that away from half filling (ie for nonzero doping x), the Curie transition is either from a metal to a metal (for not too large *g)* or from an insulator to insulator (for large *g)*; there is no insulator to metal transition unlike what is commonly seen experimentally. The effects (of temperature, magnetic field etc.) are in general smaller. Correlation U and coupling to breathing mode distortion will tend to reduce the difference in results for the doped and undoped systems.

Two approximations seem to be crucial; one is that in the adiabatic regime, ie for small $\gamma$, the static limit ($\gamma = 0$) is not only a good approximation for the physical properties, but additionally the effects of nonzero $\gamma$ are obtainable **perturbatively** (and therefore negligible for small $\gamma$) even for the large electron lattice coupling $\lambda$ present in these systems. The second is that one is not in the strong correlation (U > zt) regime, and that therefore there are no qualitative strong correlation effects. The first approximation is one of the major questions in electron phonon physics (see eg Section 5.2 below). For large $\lambda$, it is possible that the small parameter is not $\gamma$ but $\eta = \exp(-\lambda/\gamma) \ll 1$, which is not perturbative in $\gamma$. It ($\eta$) controls the effective amplitude for polaron hopping and thus the bandwidth;

for $\eta \ll 1$, the polaron is essentially site localized, whereas the conclusion from the $\gamma = 0$ limit would be that the states form a broad band with the width determined by the bare hopping broadened further by static disorder scattering. The local (*l*) polaron (and coexisting *b* or band electron) idea is developed in [56-60] and briefly described below (Section 3). A one orbital Holstein model including dynamical effects has been extensively investigated by Edwards and collaborators [61]. Early work exploring dynamical polaron effects in the Holstein model and applied to manganites is due to Roder, Zang and Bishop [62].

*2.4 Strong Hund's rule coupling $J_H$ and strong correlation U*

There is a large body of theory [63] in which the manganite is treated as an orbitally twofold degenerate electron system with strong Hund's rule coupling $J_H$ (and hence double exchange) as well as large correlation repulsion U. Ideas developed in the context of strong correlation approaches to the high $T_c$ cuprates are often used. Orbital liquid, orbital long range order, and various magnetic phases have been explored. Polaronic effects are absent by choice (essentially, *g* is assumed to be small enough to be qualitatively unimportant). There are no metal insulator transitions at general doping.

*2.5 Computer simulation*

The complexity of the manganite family and the variety of interactions (some mentioned above in Section 1) naturally suggest numerical approaches for modelling their behaviour. I do not describe this large effort, but mention one illustrative contribution, namely attempts to understand inhomogeneities on micron scales present in several manganites [5,9]. Dagotto and coworkers have argued that this is frustrated phase separation. If the system is such that one has two phases with a discontinuous transition between them such that for some parameter (say a coupling constant) they are equal or very nearly so in free energy, small fluctuations of this parameter are observed (in simulations on some model systems) to produce large regions of one phase or another. The simulations performed [64] were of frustrated square lattice Ising spin models with competing interactions, and an additional interaction chosen to produce a first order transition. Correlated disorder, eg in a random field Ising model with long range nonlocal effect of the random magnetic field, is argued to mimic the propagation of inevitable local size disorder effects in doped manganites through lattice strain. This was shown to continue to cause similar phase separation effects in higher (three) dimensions [65]. The random field Ising model itself describes two macroscopic 'states' that are degenerate in the absence of a magnetic field, and the random field mimics random local preferences for one or the other of the two. Manganites are assumed to have, for certain conditions of doping and chemical identity etc. two energetically competing phases which are however distinct enough in the nature of their order (parameter) that there is a discontinuous transition between them. The above mentioned dopant caused inevitable disorder, and its long range, elastic strain generated effect, are argued to produce sizeable domains of the two phases; the system is homogeneous however in composition or electrical charge density. If the two 'phases' are insulating and metallic respectively, and the domains are large, electrical transport can be modelled classically. A simple approach is via a resistor network (eg classical percolation theory [66]). Further, if the two regions are identified as having, respectively, antiferromagnetic and ferromagnetic coupling, its statistical mechanical simulation can lead to a strongly peaked temperature dependent resistivity. A magnetic field will affect the ferromagnetic domains substantially if these are already large, leading to a colossal reduction in resistivity. This is a mechanism for colossal magnetoresistance. An analytical Ginzburg Landau approach to the bicritical regime, and the effect of an external magnetic field in enhancing fluctuations tracked by renormalization group methods, have been discussed by Murakami and Nagaosa [67].

The proposal mentioned above for cmr may be appropriate for certain physical and chemical conditions in manganites (eg doping, ion size distribution). Indeed, experimentally, the prominent A site disorder effects in half doped manganites strongly suggest that this could be the case [6,47,48]. It seems, however, that the ubiquitous occurrence of cmr in a wide incommensurate doping range (as mentioned earlier, also [3,4,6-8,10]), requires other ideas some of which are mentioned in this review. The possible existence of more than one kind of cmr in the manganites is clearly indicated.

*2.6 The lb model*

The two $e_g$ electron per site model described by Eq.5 above is already too complicated to be investigated directly even numerically for sizeable finite systems (for an N site system with two orbital and two spin states at each site, ignoring the lattice degrees of freedom, there are $13^N$ states with upto two electrons in different orbitals at each site). In the *lb* model, one recognizes that the strong electron lattice coupling (large Jahn Teller lattice distortion) leads to the natural reorganization of the initially doubly degenerate $e_g$ orbital state at a lattice site into two states, one for which the minimum of the lattice potential energy is at $Q_o$, namely a Jahn Teller small polaron and another for which the lattice potential energy is a minimum at zero displacement; this state is therefore nonpolaronic. An electron at site i can be in one of these two states, labelled *l* and *b* respectively, with the former having a site energy lower by $2E_{JT}$ compared to the latter ($E_{JT}$ below the original degenerate level energy). Their hopping behaviour, namely kinetic energy, is very different. The *l* polaron hopping amplitude is reduced exponentially by the 'polaronic' or Huang Rhys [68] factor $\exp\{-(E_{JT}/\hbar\omega)\} = \exp\{-(\lambda/\gamma)\} = \eta \sim \exp\{-(5 \text{ to} 10)\}$; this being broadly due to the overlap between the initial and final phonon states at the site i. We argue that this exponential narrowing of the *l* bandwidth occurs for strong electron phonon coupling $\lambda > 1$ even in the adiabatic regime $\gamma < 1$, provided that $(\lambda/\gamma) >> 1$ as is the case for manganites (for these, $\lambda \sim 1$ to 5, and $\gamma \sim 0.2$ to 0.3). A calculation [69] for the one orbital per site Holstein lattice polaron model (with U=0, and using the Lang Firsov transformation [70]) estimates the leading perturbative or phonon fluctuation correction to the 'mean field' reduction factor $\eta$ to be of relative order $(t\eta/\hbar\omega_o) << 1$. There is no such reduction in the *b* electron hopping amplitude. By contrast, in theories for manganites which include the effect of strong electron lattice (Jahn-Teller) coupling, one either assumes that the classical Jahn Teller distortion results have negligible perturbative corrections because the adiabaticity parameter $\gamma$ is small [41,42], or that the quantum phonon effects act equally on both the states [71]. These characteristic strong electron phonon coupling effects are captured in a two fluid Hamiltonian $H_{lb}$, whose simplest form is

$$H_{lb} = -E_{JT}\sum_i l^+_{i\sigma}l_{i\sigma} - t\sum_{<ij>} b^+_{i\sigma}b_{i\sigma} - \mu\sum_i (n_{li\sigma} + n_{bi\sigma}) + U\sum_i n_{li\sigma}n_{bi\sigma} - J_H\sum_i (\mathbf{s}_{li} + \mathbf{s}_{bi}).\mathbf{S}_i \quad (6)$$

We notice that effects of **all** the three strong local interactions are still present in the above Hamiltonian. Effectively, the large electron lattice coupling *g* leads to two very distinct effective fermionic species or fluids, *l* and *b*; the former is a JT polaron, essentially site localized, with a lowered energy $-E_{JT}$. The latter is a band electron, hopping from site to site in a random medium which has zero site energy on hole sites and large repulsive energy U (which is composed of the Mott Hubbard correlation U* and the 'anti-Jahn Teller' state energy $E_{JT}$) on *l* polaron sites. The $e_g$ spins on site i have a strong Hund's rule ferromagnetic coupling $J_H$ with $t_{2g}$ spins at the same site. The number of electrons per site in the system is (1-x) on the average: this is the total number of *l* polarons and *b* electrons. We discuss these results in Section III, and the nature of *l* polarons and their coexistence with band *b* electrons in Section V.

Some further approximations have already been made in writing the above Hamiltonian. Firstly, we neglect the intersite hopping of *l* polarons, which is nonzero though exponentially small. In particular, we ignore the largest such term, of amplitude $\sim t\eta$ which arises from the hopping of the *l* polaron to the nearest neighbour *b* state; this hybridization is the dynamical cause of internal equilibration of the *lb* system. We assume however, that there is such an internal thermodynamic equilibrium, namely that the (1-x) electrons on the average per site are distributed among the *l* and *b* states, ie

$$\langle n_l \rangle + \langle n_b \rangle = (1-x) \quad (7)$$

with a common chemical potential μ. These constraints determine, for a given x, $E_{JT}$, t, and T, the relative number of *l* polarons and *b* electrons. One consequence of neglecting *lb* hybridization is that Eq.6 does not describe the physics of electron coherence that occurs below $T^* \sim (\eta t/k_B) \sim$ 100-150K, eg the dynamic nature of the *l* polaron, the band like behaviour of the system, the dramatic decrease of the electrical resistivity as T decreases well below $T_c$, its $T^2$ behaviour at low temperatures and the relatively small residual resistivity at T=0. We develop low energy theories here which are accurate for temperature/energy scales higher than $T^*$ and compare with a large number of experiments in this regime. Our calculations of those low temperature properties that depend on electron coherence which can develop in the metallic regime (0.2 < x < 0.4), eg electrical resistivity and Hall effect, are not reliable. On the other hand, properties such as the ground state energy are not very much affected; for example, inclusion of intersite *l* polaron hopping is expected to shift the critical concentration $x_c$ for insulator metal transition downward by at most a few per cent.

The second approximation is that though the actual intersite hopping matrix element of the *b* electron depends on the angles $\theta_i$ and $\theta_j$ describing the exact admixture of the the two $e_g$ orbital basis states, we effectively integrate out this degree of freedom, i.e. the hopping is represented by a single amplitude, which is the statistically and quantum mechanically averaged effective hopping amplitude for the *b* electron. Clearly, this is sensible in the orbital fluid regime, where there is no frozen long range order of the orbitals. In the orbital glass regime as well, the hopping amplitude t of the *b* electrons is the average over the frozen (*l* polaron) configurations (angles), and we neglect any specific effects of fluctuations with respect to this average.

The low energy sector of the *lb* Hamiltonian $H_{lb}$ (Eq.6) is the lower Hund-Hubbard band, for large $J_H$ and large U. This is the sector we focus on for comparison with appropriate physical properties. We are thus interested only in those $e_g$ (or *b* electron) states which point along the local $t_{2g}$ spin direction, and have single (or zero) occupancy on any site. Often one works in the $J_H$, U = ∞ limit so that only the lower Hund- Hubbard band has finite energy; because $J_H$ and U are large in relation to t, detailed calculations show that the $J_H$,U = ∞ limit is quite accurate. Because of this focus, in the x = 0 limit (eg the $LaMnO_3$ end) the insulating state found in the *lb* model has a large Hund-Hubbard gap; in reality there could be other unoccupied bands in this gap so that the actual activation energy for electrical conduction is smaller. In the lower Hund Hubbard band, there is a maximum of 1 *l* state and x *b* states per site.

The *l* polaron is a good excitation for energies smaller than the polaron (binding) energy $E_{JT}$. In an effective low energy theory (for energy < $E_{JT}$), one integrates out states with energy larger than $E_{JT}$. This leads to at least one new virtual double exchange interaction which is **ferromagnetic**. The interaction, second order in $t_{ij}$, is illustrated in figure 3. It arises when an *l* polaron site and a hole site are nearest neighbours i and j, and the *l* electron (**not** the polaron) hops quickly from the site i to the site j and hops back, before the lattice distortion at site i relaxes, the time scale for the latter being about $E_{JT}^{-1}$. The energy of the intermediate lattice distorted state is $2E_{JT}$. The intersite hopping matrix element depends on the angle between $t_{2g}$ spins at i and j, because of the strong onsite Hund's rule coupling between $t_{2g}$ and $e_g$ spins. For $J_H = \infty$ and classical $t_{2g}$ spins, this interaction term is, in the orbital fluid regime, given by

$$H_{VDE} \sim - (t^2/ 2E_{JT}S^2) \sum_{<ij>} \mathbf{S}_i \cdot \mathbf{S}_j [ n_i(1-n_j) + n_j(1-n_i) ] \quad (8a)$$

$$J^F = - (t^2/2E_{JT}S^2) \, x \, (1-x) \sum_{<ij>} \mathbf{S}_i \cdot \mathbf{S}_j \quad (8b)$$

where one has averaged over the orbital directions at sites i and j in Eq (8a) and further over the *l* occupations in Eq.(8b). This superexchange type of virtual double exchange coupling, second order in hopping, is ferromagnetic between the $t_{2g}$ spins because of the Hund's rule, unlike the well known Kramers Anderson superexchange for electron spins in nondegenerate orbitals which is antiferromagnetic. It involves virtual hopping, unlike the double exchange which involves real hopping and is thus linear in $t_{ij}$. We find that $J^F_{ij}$ has the right size and x dependence, and that for

0.2<x<0.4, is by far the more significant magnetic interaction than double exchange because the latter is due to the hopping of *b* electrons which are small in number in this x range. We estimate $J^F_{ij}$ to be about 2-3 meV. The interesting connection between charge (orbital) and spin degrees of freedom implied by Eq.(8a) and its intrinsically random nature have not been explored, but are quite likely connected with the A site disorder [6,47,48] and Griffiths phase [72] effects observed. Recently, several pieces of evidence supporting the above origin of intersite ferromagnetic spin coupling in manganites have turned up. From an analysis of the EXAFS lineshape in terms of the mean square spread of Mn-O bond lengths, and the temperature/doping dependence of the latter, it was inferred [73] that holes are strongly correlated with JT distorted sites as nearest neighbours; it is argued that the corresponding spins exist as ferromagnetic pairs. Considerable paramagnetic susceptibility evidence [74] points to Curie constants appropriate to nearest neighbour magnetic pairs of ferromagnetically coupled S=2 and S=(3/2) spins , ie $Mn^{3+}$ and $Mn^{4+}$.

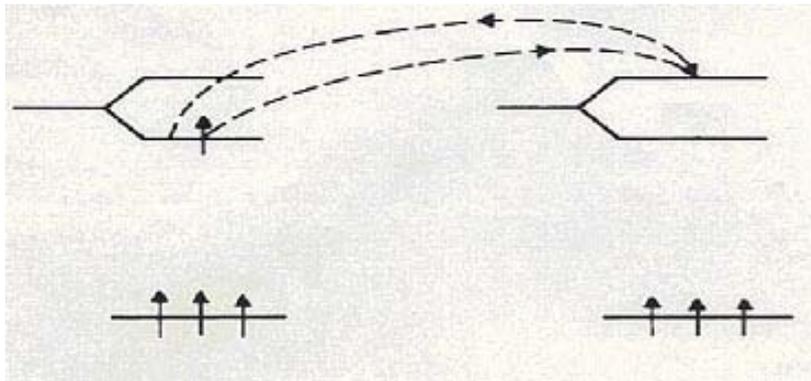

Figure 3: Schematic illustration of second order virtual nearest neighbour hopping process leading to a ferromagnetic $t_{2g}$ spin coupling.

The antiferromagnetic state of the two end members of the series, namely for x=0 and for x=1, means that there is an antiferromagnetic coupling between $t_{2g}$ spins. This is most naturally via Kramers Anderson superexchange as discussed for example in [31]. They show for example that both the size and sign of the superexchange depend on the occupied nearest neighbour (nn) $e_g$ orbital states. One can rationalize for example the occurrence of A type AF order for x=1; in the ab plane the nn $e_g$ orbitals are 'antiferro' correlated so that the $t_{2g}$ spin coupling is ferromagnetic, while in the c direction the 'ferro' correlation of the orbitals leads to an antiferromagnetic spin soupling. This is indeed observed. The nn AF coupling is of the usual form

$$J^{AF}_{ij} = - J \mathbf{S}_i \cdot \mathbf{S}_j \qquad (9)$$

## 3  Results of the strong correlation, two fluid model

I outline here some results obtained by us for the properties of manganites modelled as two coexisting fermionic fluids, one polaronic and heavy, and the other nonpolaronic and light [56-60]. As described in the previous subsection (Section 2.5) , the model assumes that the strong electron Jahn Teller phonon coupling $\lambda$ leads to a reorganization of $e_g$ states into *l* polarons with associated localized lattice distortion, and broad band *b* electrons.There is a recognition that the former have an exponentially small bandwidth because ($\lambda/\gamma$) >> 1 even though $\gamma$ < 1. A new model Hamiltonian Eq. 6 for the two *lb* fluids concretizes this dynamical separation. The relative number of *l* and *b* fermions is determined by internal equilibrium (identity of the chemical potential) and global stability (minimum free energy).

The two fluid model above is formally akin to the Falicov – Kimball model for a system of correlated f electrons and coexisting band spd electrons which was developed in the context of rare earth metals and alloys, and has been investigated extensively [75]. The presence of spin degrees of freedom of the $t_{2g}$ spins and the $e_g$ electrons is a qualitative difference.

The strong coupling approach used by us to understand the low energy behaviour of the *lb* model in the form Eq. (6) is the Dynamical Mean Field Theory (DMFT) [54]. In the single site DMFT (or self consistent impurity model) we work with, the *b* electron at a given site sees an electron bath of electrons in momentum states k with which it hybridizes (amplitude $V(\varepsilon)$ at energy $\varepsilon$). The *l* polaron has a site energy $-E_{JT}$. The local spin experiences a molecular field $\Omega$, which is the average effect of all other spins. Thus the DMFT Hamiltonian can be written as

$$H_{DMFT} = - E_{JT} n_l - \sum_{<k\sigma>} V(\varepsilon_k)[b^+_\sigma c_{k\sigma} + c^+_{k\sigma}] + \sum_{<k\sigma>} \varepsilon_k c^+_{k\sigma} c_{k\sigma} - \mu (n_l + n_b) + U n_{l\sigma}n_{b\sigma} - J_H(\mathbf{s}_l+\mathbf{s}_b).\mathbf{\Omega} \quad (10)$$

where $V(\varepsilon)$, $\mu$, and $\Omega$ ( or rather the probability $P(\Omega)$ and the resulting average magnetization m) are determined self consistently for a given temperature T, doping x and Hamiltonian parameters. Results for the *l* and *b* spectral densities at representative dopings and temperatures are shown in figure 4.

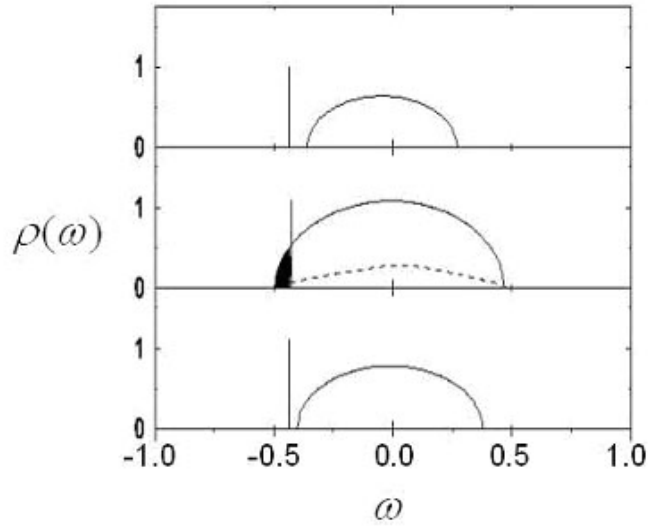

Figure 4: Spectral densities in the single site DMFT solution of the *lb* model [56,59]. The *l* level, and the *b* (band electron) spectral density (y axis) are shown in units of the bare semicircular density of states (x axis), for various dopings x and temperature T. The Jahn Teller energy $E_{JT}$ is 0.5 in these units. The model parameters are $2W = 2.4eV$, $E_{JT}=0.5eV$, $U=5eV$, and the virtual double exchange ferromagnetic coupling 2.2meV.

At T=0, the DMFT can be solved exactly [56,57,59] in the $J_H, U = \infty$ limit which is actually a very good approximation. The effective *b* bandwidth goes as $\sim \sqrt{x}$ for small x (essentially because of strong correlation U induced *b* exclusion) so that the *b* band is relatively narrow for small x. The band bottom for small x can thus lie above $-E_{JT}$ which is very nearly the chemical potential, so that there are no occupied or low energy *b* (extended)states, only localized or easily localized polaronic states are occupied, and the ground state is a ferromagnetic insulating dense polaron liquid in the DMFT (figure 4a). The system is most likely a ferromagnetic polaron glass in reality because of the pinning of *l* polarons by even weak random potentials, eg coulomb interactions, as found for example in realistic computer simulations (figure 7 and discussion there). There are also intersite orbital or polaronic correlations, explicit or induced (the latter for example via coupling to strain, or via admixture with *b* states on neighbouring sites); these can give rise to short range or even long range orbital/polaronic order. Qualitatively, the insulating ferromagnetic ground state found here for small doping (generic to manganites, but difficult to obtain in most models) is due to two reasons: the ferromagnetic virtual double exchange coupling Eq. 8 requires only that there be JT polarons and holes, not that there be moving holes as for real double exchange, so that it is quite compatible with an insulating state. Secondly, the system consists actually of only a small concentration x of holes doped into LaMnO$_3$ which is known to be a polaron crystal, so that the density of polarons can be expected to be high, about (1-x) per site; the other (*b*) electrons move in a medium with a very small concentration (x) of favourable sites (with a concentration below the percolation threshold $p_c$), avoiding polaronic sites so that their overall kinetic energy gain is smaller than the Jahn-Teller energy loss. Disorder adds crucially to this picture. Inevitable dopant disorder and the associated coulomb interaction energy can freeze or pin the *l* polarons and lead to strong nanoscopic electronic inhomogeneities, as shown in computer simulations of the *lb* model including such interactions (see below).

As doping increases, the effective bandwidth of the *b* electrons increases and there is *b* spectral density below $-E_{JT}$ (the chemical potential continues to be close to it) so that extended states are occupied and the system is a ferromagnetic metal (eg figure 4b). Thus, with increasing doping, there is a T = 0 insulator to metal transition (The critical concentration $x_c$ for this is in the observed range 0.2 for typical parameter values). This, and the insulating ferromagnetic ground state for small x described above, are two characteristic manganite phenomena which are simply understood in the two fluid model. The DMFT calculations show in detail how they depend on various material parameters.

In the metallic state, electron transport is by the mobile *b* electrons which we find to be small in density, << x. The density of JT polarons is high. They are necessarily dynamic because of eg intersite *lb* hybridization, an energetically small effect neglected in the simplest version Eq.6 of the *lb* model and in the DMFT. This process is expected to reduce their number density also selfconsistently. As temperature increases at a given doping, the effective mobile electron (*b*) bandwidth decreases because of the $t_{2g}$ spin disorder and the strong Hund's rule coupling $J_H$. Consequently, the number of occupied *b* states decreases dramatically, and one can have an insulator(figure 4c). We have also calculated the total electrical resistivity solely from the *b* electron propagators (this can be done in d=∞ where the relevant vertex corrections vanish); the *l* polarons are immobile and do not transport charge. Some results are shown in figure 5. We find a clear thermal insulator metal transition, nearly coincident with the Curie transition. Interestingly, the insulator like temperature dependence of the resistivity in the paramagnetic phase leads to an activation energy (for the *b* carrier) which is close to what is observed in systems eg LCMO and NdSrMO where such comparisons are possible [56,59].

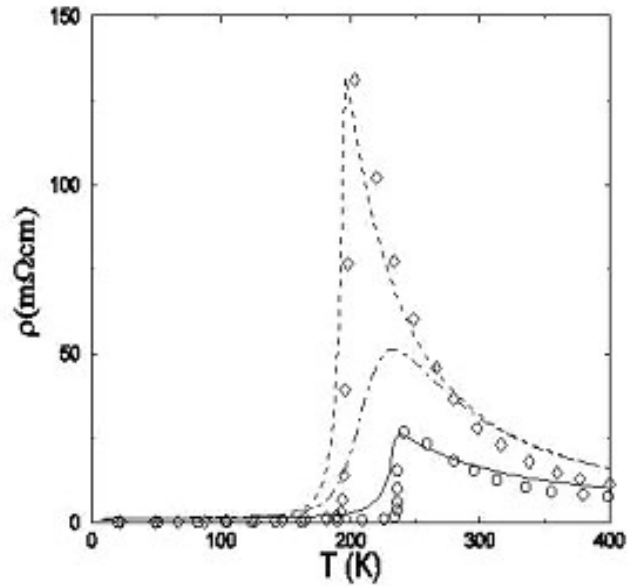

Figure 5: Resistivity (y axis) calculated in single site DMFT vs. temperature (x axis) for $E_{JT}$ = 0.5 eV, U = 5 eV and x = 0.3 [56,59]. The bare bandwidth is 2.3 eV for the dashed and dash dotted lines (H = 7T) and 2.1 eV for the full line. The diamonds are the data for NdSMO and the circles for LCMO.

Depending essentially on the ratio δ of the Jahn Teller energy $E_{JT}$ to the bare electron half bandwidth W, the thermal transition can be from insulator to insulator (large δ ) or from metal to metal (small δ) in close correspondence with experiments on manganites (figure 6).For example, the LSMO compounds are believed to have the widest bands; both the paramagnetic and the ferromagnetic states are metallic. In the narrow band system PCMO, both the phases are insulating, while in LCMO, which is believed to be intermediate, the Curie transition is from an insulator to a metal. There is strong neutron evidence that the transition is first order [76 ]; this could be because of coupling to other degrees of freedom, eg strain, neglected in the *lb* model Hamiltonian Eq.6, and in the DMFT calculation.

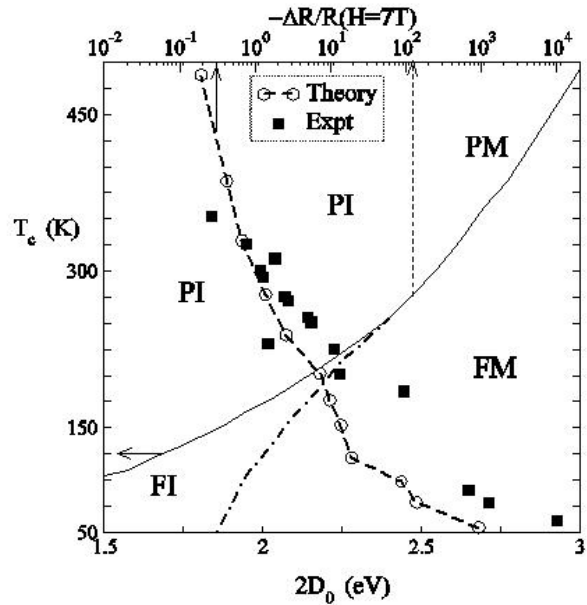

Figure 6: Materials systematics in the *lb* model [56, 59]. Ferromagnetic $T_c$ vs. bare bandwidth 2W. The full line is the calculated $T_c$ and the dash dotted line separates the calculated ferro insulator (FI) region from the ferro-metal (FM) region.The calculated and observed fractional resistance changes at 7T are also shown.

We also calculate the electrical resistivity in a magnetic field, which couples to the $t_{2g}$ and $e_g$ spins.We find a colossal decrease (figure 5). The physical reason is that the magnetic field increases the $t_{2g}$ spin magnetization; the effect is strongest near $T_c$. As a result, because of the strong Hund's rule coupling, the *b* bandwidth goes up, so that the *b* band bottom can go below the chemical potential and some extended *b* states can be occupied. The resistivity decreases dramatically. This can occur for general doping in an orbital liquid for reasons not connected with the critical value of a particular coupling constant or doping condition, bicriticality or disorder enhanced two 'phase' coexistence [9]. Ours is a generic mechanism having to do finally with the coexistence of localized polaronic and extended electronic states at a microscopic level. This can happen for strong electron phonon coupling and double degeneracy of the $e_g$ orbital. Further, in a manganite, because of the strong Hund's rule coupling, an external magnetic field can influence the conditions of coexistence of the two. The effect is strong because the mobile carrier density depends exponentially on an activation like energy which is affected directly by the magnetic field. It is present at arbitrary doping essentially because the polaron is strongly localized; the chemical potential is pinned at the polaron energy so that there is a large reservoir of electrons there.

A number of other consequences of the two fluid model have been noted [57 ]. One is that the number density of *b* electrons is much smaller than x , the hole density, and decreases strongly with temperature on a scale of $T_c$. For example, at x =0.3 in LCMO, $<n_b> \leq 0.05$ [59]. This is connected with several puzzling observations. For example, the photoemission intensity, from the earliest measurements [77] to the latest ARPES data on layered manganites [78], imply that the density of electronic states at the Fermi level is unusually small. Optical conductivity $\sigma(\omega)$ measurements [79] yield a small Drude weight; the optical sum rule oscillator strength moreover decreases strongly with increasing temperature. The plasma frequency of doped LCMO is much lower than expected [80], corresponding to a free carrier density of about 0.04. All this can be rationalized if the carrier density probed in these experiments is identified with the *b* electron density. The *l* electron has an exponentially small quasiparticle weight at the Fermi energy; removal of the electron of a polaron by the photon actually shows up as a mid-infrared peak, centred at the polaron binding energy, whose contribution is excluded from the Drude weight. A comparison of $<n_b>$ with the Drude weight in LSMO [79] is quantitatively successful both in absolute value and in temperature dependence, for a set of model *lb* parameters chosen to reproduce some other properties of the system (eg $T_c$, $\rho(T_c)$).[2] The large isotope effect on $T_c$ can be rationalized in terms of the exponential dependence of the l polaron bandwidth reduction factor $\eta$ on the inverse square root of the isotope mass. The *l* polaron contribution to $T_c$ via double exchange is directly proportional to this bandwidth.

The *lb* model has been extended [82] to include electrostatic coulomb interaction in the doped mixed valent manganites. It is apparent that in its absence, the *l* polarons and the *b* electrons will phase separate, since then the *b* electrons gain maximum kinetic energy. However, including coulomb interactions and imposing the equilibrium condition of the equality of **electro**chemical potential, one finds that macroscopic phase separation is muted to one of nanoscopic inhomogeneity with puddles of hole regions (home for *b* electrons) and regions of immobile *l* polarons. The scale of the inhomogeneity is set by the randomness of the dopant ions and the strength of coulomb interactions setting a screening length scale of a few nanometers in the low doping region. Figure 7 shows two examples of numerical simulation of the *lb* model Eq.6 with the effect of additional coulomb interactions included in the Hartree approximation, for a $16^3$ system with doping x=0.3 and x=0.4. The simulation is of the ground state for $J_H, U = \infty$. It is clear that the system is very inhomogeneous on a microscopic scale, though single phase on large length scales in the thermodynamic sense (insulating for x=0.3 and percolatively metallic for x=0.4 in figure 7). The intrinsic coulomb disorder and consequent pinning are connected with the absence of heavy fermion effects in the hybridized *lb* system, unlike rare earth intermetallics for example. The insulating regions are polaronic; if intersite correlations between them are included, the charge inhomogeneities do not change but short range 'antiferro'-order naturally develops and the picture is very close to what is seen in STM experiments. Coulomb interaction as the cause of electronic inhomogeneities has been discussed for long, eg in connection with frustrated phase separation in the t-J model for cuprates [83], and in the context of the low density 2d electron gas (droplet phases, [21]). It has been recently suggested [22] that in strongly correlated systems, two different phases (with different kind of long range order ?) are likely to be close in energy; this is termed electronic softness. In such a case, ubiquitous long range coulomb or strain related interaction can lead to strongly inhomogeneous states. While this scenario is plausible, our simulation addresses the question of electronic inhomogeneities microscopically, in a manifestly strongly correlated system. The two states are **not** distinguished by different kinds of order, but are quantum mechanically distinguished, by local energy and dynamics. In the presence of strong correlation (U in our case), there is a tendency for phase separation, which is restricted to nanoscopic scales by coulomb interactions. The micron scale inhomogeneities seen in a number of systems, eg via electron microscopy [15 ] or via position sensitive ARPES [15], are most likely due to other, elastic strain related causes.

---

[2] There are more recent measurements of $\sigma(\omega)$ which do not show a low frequency Lorentzian or simple Drude form [81]. This could finally be due to *l* polaron coherence developing in these systems in the metallic phase as temperature decreases. The effect of *lb* hybridization on *b* electrons, via a strongly temperature and energy dependent scattering, as well as absorption by dynamic polarons, is quite likely to change $\sigma(\omega)$ qualitatively. The earlier measurements, eg [79], could be under conditions such that because of pinning, *l* polaron coherence is absent in them.

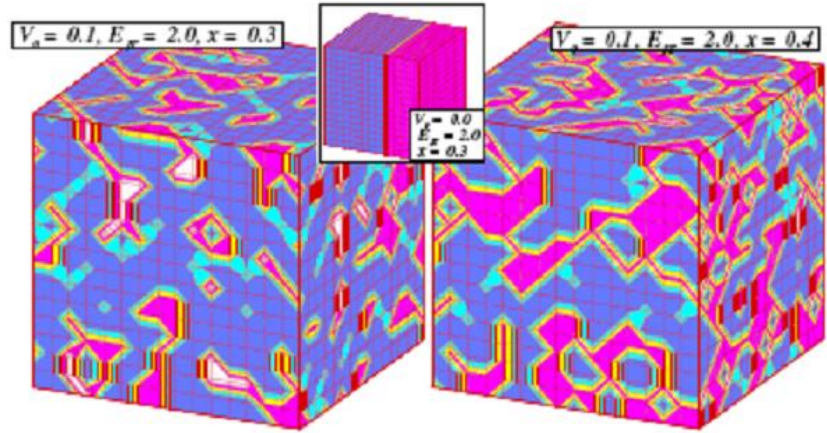

Figure 7: (colour online) Real space electronic distribution obtained from simulations on a $16^3$ cube. Magenta (darkest) denotes hole clumps with occupied *b* electrons, white (lightest) denotes hole clumps with no *b* electrons, cyan (2$^{nd}$ lightest) denote singleton holes, and light blue (2$^{nd}$ darkest) represents regions with *l* polarons. The configuration on the left shows isolated clumps with occupied *b* electrons (*b* puddles). For larger doping, percolating clumps are obtained and the system is a metal (right). The inset shows results in the absence of long range Coulomb interaction and shows macroscopic phase separation. All energy scales are in units of t.

## 4  Current issues

Several open areas in manganite physics are also subjects of a great deal of current activity. Some have already been mentioned in earlier parts of this review. Examples are the role of A site disorder in half doped manganites (Section 1.2), the possibility of different kinds of colossal magnetoresistance (Section 2.5, Section 3) and the overarching question of 'phase' separation (Section 1.1).An area which is seeing high quality experimental activity is that of ARPES experiments possible in bilayer manganites [78]. These show a pseudogap in the electronic density of states near the Fermi energy, resembling (in size, energy location, and most interestingly in angular dependence in the plane) that seen and well documented in underdoped cuprates.

## 5  Some questions arising

I conclude this review by mentioning two questions possibly of broad interest raised by observations in manganites, both having to do with new physics that probably arises for strong electron phonon coupling as doubtless present in manganites. The first is whether the strong electron phonon coupling effects in them can be usefully described in terms of polarons, and if so of what kind? The second , related question, is the physics of strong electron phonon coupling in adiabatic polaronic systems, namely systems in which $\lambda > 1$ and $\gamma < 1$, but $(\lambda/\gamma) \gg 1$.

5.1 *Polarons :*

A large number of phenomena in manganites indicate the existence of polarons in them over a wide range of conditions; these composites of an electron and associated lattice distortion affect their physical properties fundamentally. In this subsection, I mention some of these phenomena and properties. Many of these have been discussed in terms of existing models for polarons, largely following the early work of Holstein [84] on an orbitally nondegenerate electron at a lattice site inducing displacement of an Einstein mode there. Whether one has polarons of this kind or others more specific to the local symmetry and interactions in manganites, theirnumber, correlation and dynamics, their role in the observed phenomena, and their direct observability are some of the major questions in the field as mentioned below.

As pointed out earlier, the twofold degeneracy of the $e_g$ orbital of Mn is naturally removed by the Jahn Teller effect, namely by breaking of its local octahedral symmetry since the coupling $g$ between the $e_g$ electron onsite and the $\mathbf{Q_i}$ modes (Eq. 3a) is large ($g \sim$ 2eV/A). The composite of the $e_g$ electron and the associated local lattice distortion is the Jahn Teller polaron, widely believed to be present and most likely significant in manganites. (The local excess /deficit d electron density also induces a large breathing mode or uniform distortion of the octahedron of which the Mn ion is the centre; its consequences are not discussed here). I describe some observations argued to attest to existence of such polarons and theories including the *l* polaron *b* electron (two fluid) model ([56-60] and Sections II and III above) used to describe phenomena in these systems.

Some relatively direct evidence for polarons in manganites follows from 'instantaneous' or short time scattering experiments using pulsed high energy neutrons or EXAFS, which determine pair distribution functions or PDF's [16-18]. The results show that in doped manganites, short and long Mn-O bond lengths exist on timescales over which the high energy neutron or the X-ray traverses atomic (nanoscale) regions. The bimodal bond length distribution, most naturally interpreted as an appropriate Jahn Teller distortion, is observed not only at or very close to zero doping, or at high temperatures, or in the insulating phase, but even in the most metallic of manganites, namely

$La_{1-x}Sr_xMnO_3$, into the metallic regime, for $x>x_c$ and $T<T_c$. Signatures of local distortion are observed for conditions of doping and temperature such that there is no long range lattice distortion generated superstructure (unlike in $LaMnO_3$ below 780K); the system is pseudocubic. The decrease in the intensity of the bimodal distribution and in the bimodal dispersion with increasing x and with decreasing T [16,18] suggests several possibilities: there are fewer Jahn Teller polarons as x increases, the polarons become dynamic progressively with decreasing T and increasing x, and there is a decrease in the polaronic distortion. All these are likely to feed back on each other. In neutron scattering experiments, observations of diffuse scattering above $T_c$ [85] have been described in terms of a polaron glass, i.e. a static arrangement of local J-T polarons. At lower temperatures (eg below $T_c$, in the ferromagnetic metallic regime) this scattering disappears; the local polaron lifetime becomes smaller than the (thermal) neutron traversal time. Observations of the mid-infrared peak in optical conductivity $\sigma(\omega)$ [86], of nondispersive states about an eV or so below the Fermi energy in ARPES [87], of local cubic symmetry forbidden JT phonon lines in light scattering [37] have all been interpreted in terms of the presence of JT polarons. An assumption underlying these suggestions is that the conditions of small JT polaron formation are local (on site), and are not therefore much affected (adversely) by doping and possible metallicity which gives rise to electrostatic screening. In the manganites, screening is poor, and the screening length scales are indeed larger than the unit cell size relevant for the kind of JT polaron formation mentioned above. However, polaronic signatures are feeble in the metallic, lower T regime.

One simple well known model is the Holstein polaron [84], which arises when a tight binding electron in an orbitally nondegenerate state on a lattice site couples strongly to a local Einstein phonon mode. It is well known that in this model, the ground state for a single electron is extended for $g<g_c$ and is a self trapped (Holstein) polaron for $g > g_c$, with a rapid crossover. It is assumed that the many electron system will consist of either band electrons or of polarons, depending on $g$. In the latter case, it is argued (mostly in the antiadiabatic regime) that the polarons form an exponentially narrow band at low T, and at high temperatures hop incoherently from one site to another, surmounting (via coupling to thermal 'noise') the energy barrier associated with intermediate lattice displacement configurations. It has been argued, eg[4] that transport properties of manganites in the orbital fluid regime (approximately 0.2<x<0.4), eg electrical resistivity ρ, Hall coefficient $R_H$ and thermopower can be understood this way. While Holstein polarons have been widely discussed [4] there is no investigation of the transferability of results for them to the rather different situation of manganites in which the tight binding $e_g$ electrons which can form polarons are orbitally twofold degenerate, can have high density, and other strong local correlations/interactions are present.

The other generic polaron model, namely the Frohlich model, arose for ionic solids, where an added electron produces long range dipolar electrostatic distortion of the lattice. The importance of such polarons for manganites has been argued for by Alexandrov and coworkers [88].

As mentioned earlier above, Millis and coworkers [41,42] pointed out the importance of local electron phonon coupling in the context of manganites, namely for electrons in doubly degenerate $e_g$ states. (They took the phonons to be classical, i.e. treated the lattice displacements as classical variables, with the statistical distribution of the JT distortion assumed to be the same at each site and determined self consistently). Because the phonons are treated classically, and the system has annealed disorder, the $e_g$ electrons move in a static random medium. The polaron states form a band, further broadened by disorder. Millis et al found that the the average polaronic distortion can be sizeable at high temperatures and decreases with cooling. The low energy electronic states are organized into two lower Hund bands, one polaronic and another not, somewhat like the *l* and *b* states discussed above. The polaronic band is broader than the *b* band because of static disorder effects. In our two fluid model the *l* polarons are also composite excitations of the orbitally twofold degenerate $e_g$ electrons and JT lattice distortions. However, they have an exponentially small bandwidth forming an essentially sharp energy level and are dynamically very distinct from *b* electrons being much slower and much more easily localizeable. Because of this difference in the nature of the energy distribution of the JT polaronic states, in the *lb* model for example, the density of polarons decreases only a little with decreasing temperature in the orbital fluid regime in contrast to the expectation in the static polaron model of Millis et al.. The difference arises from our argument that even though the adiabaticity parameter γ (=ℏω/t) is much less than unity (0.2 to 0.3) in manganites, one is in the exponentially narrowed polaron bandwidth regime, with the narrowing factor being exp(-λ/γ) << 1. This is in contrast to the well studied perturbative Migdal regime [89] where λ needs to be <1 which is not the case for manganites. The general question of the physics of strong electron phonon coupling is outlined below.

The *l* polaron *b* electron model described above differs from the polaron models used so far for manganites in several significant ways, some already mentioned. Firstly, because of the twofold degeneracy of the $e_g$ level, a two fluid description is natural. The *l* polarons which can form a dense liquid or glass or crystal , are site localized or nearly so. The electrical transport is primarily by the *b* electrons, though in principle by both *l* polarons and *b* electrons. Thus, metallic conduction can coexist with the presence of polarons and will occur when there is a sizeable density of *b* electrons in the system. In the insulating phase, at high temperatures for instance, when the density of *b* electrons is low due essentially to the small probability for thermal activation, diffusion (or variable range hopping) of *l* polarons can contribute significantly to electrical transport. The density and nature (eg static/dynamic, size) of the *l* polaron is expected to change significantly with temperature, doping and chemical identity of the manganite. In this picture, nanoscopic inhomogeneities (of the *l* and *b* regions) are inevitable, and are due to coulomb interactions.

5.2 *Electron Phonon Coupling:*
An unresolved question of great basic and current significance insistently thrown up by experimental results on oxides is the physics of large electron phonon coupling in them, broadly expected because of the relatively localized nature of the unfilled shell d states abetted by electron correlation. Its strength can be characterized by a parameter *g* in a schematic Holstein like electron lattice Hamiltonian

$$H_{Holstein} = -t \sum_{<ij>} b^+_i b_j + g \sum_i n_i x_i + (1/2) \sum_i (K x_i^2 + M^{-1} p_i^2) \qquad (11)$$

where $x_i$ is the displacement of a one dimensional Einstein mode (frequency $\omega_o = \sqrt{K/M}$) at site i, and $n_i$ is the electron number at site i. The form for a doubly degenerate orbital has been described earlier $H_{JT}$ (Eq.3a). This also is characterized by a single coupling constant $g$. The dimensionless parameter $\lambda = (g^2/Kt)$ is a measure of its strength vis a vis the lattice displacement energy $(1/2)K x_i^2$ and the electron kinetic energy (nearest neighbour hopping amplitude t, bare bandwidth 2W = 2zt for a tight binding band with z nearest neighbours). Another dimensionless parameter of relevance is the adiabaticity or phonon energy relative to the electron energy, namely $\gamma = (\hbar \omega_o/t)$. For a free electron gas of Fermi energy $\varepsilon_F$, the parameter is ($\hbar \omega_o/\varepsilon_F$). In manganites, given the range of values for g, K, t, and $\omega_o$, one has $\lambda \sim 1-4$ and $\gamma \sim 0.2$ to 0.3. Thus one is in the strong electron phonon coupling but adiabatic regime.

Figure 8 shows the various regions in the electron phonon coupling – adiabaticity ($\lambda, \gamma$) plane. The best known is the Migdal [89] or the weak coupling-adiabatic region, for which one has $\lambda < 1$, and $\gamma << 1$, i.e. the free Fermi gas or a Fermi liquid with fast electrons coupled weakly to slow lattice vibrations. Higher order corrections to the electron phonon vertex, basically the electron wavefunction modification, are small and weakly perturbative, of relative order $\lambda\gamma$ and higher (since $\lambda<1$, this translates into $\gamma<< 1$). It has been realized for long [90] and has been recently demonstrated [91,92] through detailed calculations in the model described by Eq. (11), that there is an instability for $\lambda = \lambda_c \sim 1$. Migdal theory breaks down. Engelsberg and Schrieffer [93] performed in the sixties a detailed calculation for a coupled continuum electron phonon system with free particle dispersion for the unperturbed electrons, an Einstein phonon of bare frequency $\omega_o$ and the electron phonon coupling Eq. (11), and concluded that for large $\lambda$ there is no convergent perturbative expansion. The physical reason is well known; for large $\lambda$, the state with an electron bound by the lattice distortion it causes, namely a small polaron, has lower energy. The zeroeth order electron state wavefunction is not a simple plane wave, and the phonon wavefunction is that of a **displaced** harmonic oscillator. Perturbative description about the lattice undistorted ground state is therefore inappropriate, and such theories are not expected to be convergent. This is most easily reflected in the large electron phonon coupling – antiadiabatic or large ($\lambda,\gamma$) region of figure 8. Here the site localized electron forms a small polaron and the effective intersite hopping is the bare one multiplied by the (initial and final) phonon wavefunction overlap or Franck Condon factor [68] which is exponentially small: $t^*_{ij} \sim t_{ij} \exp(-\lambda/\gamma) = t_{ij}$. No systematic small perturbative expansion parameter for this regime analogous to $\lambda\alpha$ in the Migdal regime, is well known. There can be a small number, namely $\eta = \exp(-\lambda/\gamma)$, but this is not analytic in the possible small parameter ($\gamma$).

The manganites (figure 8) are in the regime $\lambda > 1, \gamma < 1$, and $(\lambda/\gamma) >> 1$. In this adiabatic strong coupling regime it is often assumed quite plausibly that though the ground state is or can be polaronic, the corrections due to the small adiabaticity factor $\gamma$ are perturbatively small [41,42]. However, we have taken the view above (Sections II and III) that in this polaronic regime as well, the exponentially small factor $\eta = \exp(-\lambda/\gamma)$ is operative. It reduces the polaron hopping drastically, and a natural scheme incorporating this effect is the two fluid (localized polaron-band electron) or *lb* model. The parameter $\eta$ has an essential singularity at $\gamma = 0$, namely there is no perturbative expansion around the static limit, even for extreme adiabaticity ($\gamma << 1$). A simple two site one electron model [94,95] with electron phonon coupling of the form Eq. (11) bears this out. In the adiabatic (Born-Oppenheimer) limit $\gamma \to 0$, on solving for the eigenvalues of the Hamiltonian including hopping, one finds that for $\lambda > \lambda_c \sim 1$ there is a twofold degenerate polaronic minimum for the lowest energy eigenvalue. The kinetic energy term lifts this degeneracy through ion quantum tunneling. The splitting is exponentially small, approximately $t_{12} \exp(-\lambda/\gamma)$ and is thus perturbatively inaccessible for small $\gamma$. A similar result has been obtained for the half filled Holstein model [96] in the unbroken symmetry phase. Whether this exponentially small polaron energy splitting (appropriately generalized) survives for similar reasons and for arbitrary filling, the effects of orbital degeneracy, of admixture with
nonpolaronic broad band states, of a finite density of polarons, of proximity to commensurate densities, of dimensionality and lattice coordination, are questions being explored.

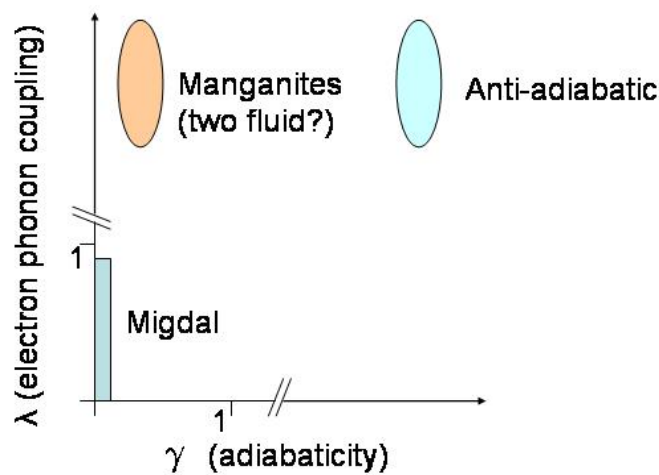

Figure 8: Regions in the electron phonon coupling ($\lambda$) adiabaticity ($\gamma$) plane. The Migdal, anti-adiabatic and manganite regimes are shown.

## 6  Acknowledgements

I would like to thank the many colleagues and students at the Indian Institute of Science, Bangalore for introducing me to this fascinating field, and for continuing discussions. Some of them are C N R Rao, A K Raychaudhuri, A K Sood, S V Bhat, H R Krishnamurthy, V K Shenoy, G Venketeswara Pai and S R Hassan. I would also like to thank Dr Arindam Chakraborti for crucial help with the manuscript.